\documentclass[aps,twocolumn,footinbib,11]{revtex4}
\usepackage{amsmath}
\usepackage{epsfig}
\usepackage{subfigure}
\usepackage{float}
\usepackage{longtable}
\usepackage{times}
\usepackage{txfonts}
\usepackage{color}
\usepackage{ulem}

 \usepackage{ifpdf}
 \ifpdf
 \usepackage[pdftex]{hyperref}
 \else
 \usepackage[hypertex]{hyperref}
 \fi

 \hypersetup{
   pdftitle={},%
   pdfauthor={},%
   pdfsubject={},%
   pdfkeywords={},%
   pdfstartview={},%
   bookmarksopen=true, breaklinks=true, debug=true, %
   colorlinks=true, linkcolor=red, citecolor=blue, urlcolor=blue
 }

\newcommand{\beq}{\begin{eqnarray}}
\newcommand{\eeq}{\end{eqnarray}}
\newcommand{\be}{\begin{eqnarray*}}
\newcommand{\ee}{\end{eqnarray*}}

\newcommand{\eqs}[1]{\begin{equation} \begin{split} #1\end{split} \end{equation} }

\newcommand{\ie}{{\it i.e.}}
\newcommand{\etal}{{\it et al.}}
\newcommand{\eg}{{\it e.g.}}

\newcommand{\ce}[1]{Eq.~\eqref{#1}}

\newcommand{\cf}[1]{{Fig.~\ref{#1}}}
\newcommand{\ct}[1]{{Table~\ref{#1}}}

\def\lsim{\raise0.3ex\hbox{$<$\kern-0.75em\raise-1.1ex\hbox{$\sim$}}}
\def\gsim{\raise0.3ex\hbox{$>$\kern-0.75em\raise-1.1ex\hbox{$\sim$}}}

\def\pPb  {$p$Pb}
\def\pPbm  {p\mathrm{Pb}}

\def\Ncoll   {\mbox{$N_{\rm coll}$}}

\def\RpPb    {\mbox{$R_{p\rm Pb}$}}

\def\jpsi    {\mbox{$J/\psi$}}

\def\beq     {\begin{equation}}
\def\eeq     {\end{equation}}

\long\def\symbolfootnote[#1]#2{\begingroup%
  \def\thefootnote{\fnsymbol{footnote}}\footnote[#1]{#2}\endgroup}

\begin{document}


\title{Impact of the Nuclear Modification of the Gluon Densities 
on $\boldsymbol\jpsi$ production in $\boldsymbol{p}$Pb collisions at $\boldsymbol{\sqrt{s_{NN}}=5}$ TeV}

\author
{E. G. Ferreiro$^a$, F. Fleuret$^b$, J.P. Lansberg$^c$, A. Rakotozafindrabe$^d$}

\affiliation{
$^a$ Departamento de F{\'\i}sica de Part{\'\i}culas, Universidade de Santiago de Compostela, 15782 Santiago de Compostela, Spain\\
$^b$Laboratoire Leprince Ringuet, \'Ecole polytechnique, CNRS/IN2P3,  91128 Palaiseau, France\\
$^c$IPNO, Universit\'e Paris-Sud, CNRS/IN2P3, F-91406, Orsay, France \\
$^e$IRFU/SPhN, CEA Saclay, 91191 Gif-sur-Yvette Cedex, France}

\begin{abstract}
We update our previous studies of nuclear-matter effects on \jpsi\ 
production in proton-nucleus for the recent
LHC \pPb\ runs at  $\sqrt{s_{NN}}=5$ TeV.  
We have analysed the effects of the modification of the gluon PDFs in nucleus, using 
 an exact kinematics for a $2\to 2$ process, namely $g+g\to J/\psi+g$ as expected from LO pQCD.
This allows us to constrain the transverse-momentum while computing the nuclear modification factor for different rapidities, 
unlike with the usual simplified kinematics. 
Owing to the absence of measurement in $pp$ collisions at the same $\sqrt{s_{NN}}$
and owing to the expected significant uncertainties in yield interpolations which would hinder definite interpretations of
nuclear modification factor --$R_{\pPbm}$--, we have derived forward-to-backward and central-to-peripheral yield ratios in which the unknown proton-proton yield cancel.
These have been computed without and with a transverse-momentum cut, \eg~to comply 
with the ATLAS and CMS constraints in the central-rapidity region.
\end{abstract}

\maketitle




{\it Introduction.---} In this brief report, we proceed to an update  
of our earlier studies~\cite{Ferreiro:2008wc} of the 
nuclear-matter effects on the production of $J/\psi$ in proton-nucleus at RHIC
for the LHC experimental conditions. 
 In these previous studies, we have indeed  shown that the way to accurately evaluate
the parton kinematics --and thus the nuclear shadowing-- depends on the $J/\psi$ 
partonic-production mechanism.  
Doing so, we could go beyond other $J/\psi$-production studies in $pA$ collisions~\cite{OtherShadowingRefs} 
in which it was assumed 
that the $c\bar{c}$ pair was produced by a \mbox{$2\to 1$} partonic process where the 
colliding gluons necessarily carried intrinsic transverse momentum~$k_T$, entirely transferred 
to the quarkonium final state. 

Our earlier works --as well as this update-- account for a
kinematics corresponding to a \mbox{$2\to 2$} partonic process for $J/\psi$ production at $\alpha_s^3$,
as in the Colour-Singlet Model (CSM) or the Colour-Octet Mechanism (COM). In both cases, the partonic process is
similar, \ie~$gg\to J/\psi +g$.

Our initial motivation stemed from the recent findings that $J/\psi$ production at low $P_T$ 
--where most of the $J/\psi$'s are-- likely proceeds via colour singlet 
transitions~\cite{Brodsky:2009cf}. Recent studies 
of  QCD corrections have  confirmed this by showing, 
on one hand, that the too soft $P_T$ dependence of the LO 
CSM~\cite{CSM_hadron} was significantly improved when incorporating $\alpha_S^4$ and 
$\alpha_S^5$~\cite{QCDcorrections} topologies 
and, on the other,  that the yield predicted by the NLO CSM for $e^+e^-\to J/\psi+X_{{\rm non}\ c \bar c}$~\cite{Gong:2009kp,Ma:2008gq} saturates the the Belle experimental values~\cite{Pakhlov:2009nj}. 
The COM component~\cite{Zhang:2009ym}, which happens to be precisely the one appearing 
in the low-$P_T$ description of hadroproduction via a $2\to 1$ process~\cite{Cho:1995ce}, 
is therefore likely not significant. By the way, this confirms the results global survey 
of low energy data by Maltoni \etal~\cite{Maltoni:2006yp} and the recent one by 
Y. Feng~\etal~\cite{Yu:2013}. Yet, as we mentionned in the previous paragraph, a $ 2\to 2$ kinematics is also 
the relevant one in the COM for finite $P_T$ (see \eg~\cite{Sharma:2012dy}).
 
To follow the lines of our earlier works, we have used a generic \mbox{$2\to 2$} matrix element
which matches the $P_T$ dependence of the LHC data (dubbed  
{\it extrinsic} scheme in~\cite{Ferreiro:2008wc}) such as $g+g \to J/\psi + g$, as opposed 
to a \mbox{$2\to 1$} process as it can be in the colour evaporation model  and colour octet 
mechanism at low $P_T$ (for recent reviews see~\cite{Lansberg:2006dh,Brambilla:2010cs}).

We therefore present here our results for proton-lead collisions at 5~TeV.  We have studied the effect 
nuclear modification of the gluon PDFs on the $J/\psi$ yield and as illustration of possible additional effects that
of the so-called effective absorption --the break-up of the pre-resonant pair along its way off the lead nucleus--.

Other phenomena as such the Cronin effect, coherent power corrections and nuclear-matter-induced 
energy loss may also be at work. The latter effect is the subject of recent active 
activities~\cite{Vitev:2007ve,Arleo:2012hn} whose conclusions may seem contradictory. As 
such and since this work has only as ambition to provide updated predictions at LHC 
energies based on~\cite{Ferreiro:2008wc}, we have not considered these effects.

{\it Theoretical framework.---} We have thus used our probabilistic Glauber Monte-Carlo framework, 
{\sf JIN}~\cite{Ferreiro:2008qj}, 
which allows us to encode $2\to 2$ partonic mechanisms for $J/\psi$ production.
As far as the nuclear modification of the gluon PDF is concerned, we have employed the parametrisation 
EPS09~\cite{Eskola:2009uj} and nDSg~\cite{deFlorian:2003qf}. The former comes 
with several sets to be used to map out the nPDF fit uncertainties. The spatial 
dependence of the nPDF  has been included in our approach, assuming an inhomogeneous shadowing 
proportional to the local density~\cite{Klein:2003dj,Vogt:2004dh}. The behaviour of the 
different sets we have used is depicted on \cf{fig:R_g^Pb}. We note that, for instance, 
the range spanned by the nCTEQ~\cite{nCTEQ} and DSSZ~\cite{deFlorian:2011fp} parametrisations can even be wider. See also~\cite{Frankfurt:2011cs}
for a recent review.

Because of the much larger Lorentz boost between the lead rest frame compared to RHIC energies, the heavy-quark
pair propagating in the nuclear matter will nearly always be in a pre-resonnant state, \ie~smaller than the physical 
$J/\psi$ size. Indeed, the relevant timescale to analyse the pair evolution is its formation time, 
which, following the uncertainty principle, is related to the time needed -- in their rest frame -- to 
distinguish the energy levels of the $1S$ and $2S$ states,
$t_f=  \frac{2 M_{c\bar c}}{(M^2_{2S}-M^2_{1S})}= 2 \times 3.3$ GeV / 4 GeV$^2= 0.35$ fm 
for the $\psi$. $t_f$ has to be considered in the rest frame of the target nucleus.
\ct{tab:tf-LHC} summarises the various formation times in the lead rest frame. Only in 
the most backward region does $t_f$ get to the order of the lead radius, $r_{\rm Pb}$. 

The probability for the heavy-quark pair to be broken up during 
its propagation through the nuclear medium is 
commonly known as the nuclear absorption, usually 
parametrised by an effective break-up cross section~$\sigma_{\mathrm{eff}}$.
Since the pair is smaller (${\cal O}(1/(2m_c))$ instead of ${\cal O}(1/(\alpha_s(2m_c)2m_c))$, 
we have considered smaller --1.5 and 2.8 mb-- values for the 
effective break-up cross section. Note that a pre-resonant pair along its way off the 
lead nucleus can still be broken up if, for instance, the scatterings increase its invariant 
mass until above the open-charm threshold (see~\eg~\cite{Qiu:1998rz}). That being said, our
result with nonzero break-up cross section are primarily shown for illustrative purposes. Only
qualitiative statements should be drawn from them.

\begin{table}[tb!]
\begin{center}\setlength{\arrayrulewidth}{1pt} \footnotesize
\begin{tabular}{ccc|ccc|ccc}
\hline\hline
 $y$ & $\gamma(y)$ & $t_f(y)$    &  $y$  & $\gamma(y)$ & $t_f(y)$  &  $y$  & $\gamma(y)$ & $t_f(y)$\\
\hline
-4.0 & 20     & 6    fm &      -0.5 & $10^3$            & $3.0\times 10^2$   fm    &   1.5 & $6\times 10^3$   & $2.3\times 10^3$  fm \\
-3.5 & 50     & 15   fm &       0.0 & $1.7\times 10^3$  & $5.3\times 10^2$   fm    &   2.5 & $1.5\times 10^4$   & $6.0\times 10^3$ fm \\
-2.5 & 140    & 45   fm &       0.5 & $2.7\times 10^3$  & $8\times 10^2$ fm    &   3.5 & $4.2\times 10^4$   & $1.7\times 10^4$ fm \\
-1.5 & 370    & 110  fm &           &                   &                        &   4.5 & $1.2\times 10^5$   & $6.0\times 10^4$ fm \\
\hline\hline
\end{tabular}
\caption{Boost and formation time in the Pb rest frame of the $J/\psi$
at $\sqrt{s_{NN}}=5$ TeV in $p$Pb collisions ($E_N^{\rm Pb}=1.57$ TeV and the Pb has a negative rapidity).}\label{tab:tf-LHC}
\end{center}\vspace*{-.75cm}
\end{table}

\begin{figure}[htb!]
\includegraphics[trim = 0mm 0mm 0mm 0mm, clip,width=\columnwidth]{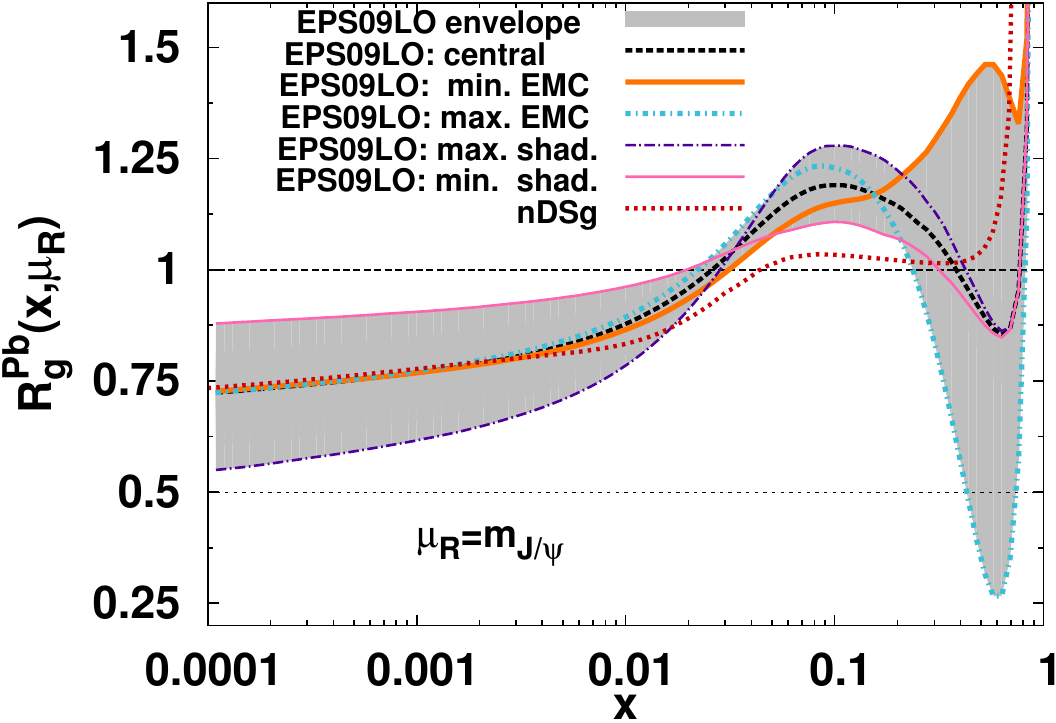}
\caption{(Color online) The nuclear modification of the gluon density in a lead nucleus at a scale taken to be 
the $J/\psi$ mass as obtained with the nDSg set, the envelope of the 30 EPS09 LO sets as well as its central set and 4 specific sets reproducing its 
envelope for the gluon.}
\label{fig:R_g^Pb}\vspace*{-.25cm}
\end{figure}

{\it Relevant observables for the 2013 LHC proton-lead run.---}
One usually characterises the suppression of the \jpsi\ by the {\it nuclear modification factor}, 
$R_{pA}$,
the ratio of the \jpsi\ yield in $pA$ collisions to the \jpsi\ yield in 
$pp$ collisions at the same energy multiplied by the average
number of binary collisions in the proton-nucleus, $\langle \Ncoll\rangle$:
\beq
R_{pA}=\frac{dN_{pA}^{J/\psi}}{\langle\Ncoll\rangle dN_{pp}^{J/\psi}}.
\eeq
Any nuclear effect affecting $J/\psi$ production leads to a deviation of $R_{pA}$ from {\it unity}.

Yet, in the absence of a yield measurement at the same energy, $N_{pp}^{J/\psi}$, the normalisation of such 
a factor depends on an interpolation which brings in additional systematical uncertainties. If the nuclear 
modifications
are of the order 10 \% in a given kinematical region, it is therefore likely that the measured nuclear modification 
factor will not be precise enough to call for a yield suppression or a yield enhancement. The same 
applies for comparison with theoretical calculations.

In such a case, to keep a specific character of an {\it experimental measurement} showing {\it unity}, 
one can resort to two additional ratios 
emphasising the rapidity or the centrality dependence of the nuclear effects. Forward-to-backward ratios 
can then be formed such
\eqs{R_{\rm FB}(|y_{CM}|)\equiv \frac{dN_{pA}^{J/\psi}(y_{CM})}{dN_{pA}^{J/\psi}(-y_{CM})}=\frac{R_{p{\rm Pb}}(y_{CM})}{R_{p{\rm Pb}}(-y_{CM})}
\label{eq:RFB},} 
in given rapidity and/or $P_T$ bins. Since
the yield in $pp$ is symmetric in $y_{CM}$, it cancels out in the double ratio in the l.h.s. of \ce{eq:RFB}.
Central-to-peripheral ratios, for instance
\begin{eqnarray}
  \label{eq:rcp}
  R_{\rm CP} = \frac{\left(\frac{dN_{J/\psi}}{dy}/\langle \Ncoll \rangle
    \right)}{\left(\frac{dN^{60-80\%}_{J/\psi}}{dy}/\langle N^{60-80\%}_{\rm coll}\rangle, 
    \right)}.
\end{eqnarray}
or conversely peripheral-to-central ratios, are more common and they are recognised
to reduce  experimental systematical uncertainties, such as the luminosity as well as 
acceptance and efficiency  corrections. A value close to unity at 
least indicates the absence of a centrality dependence of the nuclear effects.

{\it Results.---} Before showing the results for $R_{\rm FB}$ and $R_{\rm CP}$ which can directly be compared to data, 
we have found it important to emphasise three features of our theoretical predictions --likely also 
pertaining to other works-- which may be overlooked along the way of the comparisons between experimental
and theoretical results. Indeed, we currently have at our disposal improved nPDF fits with error analysis, 
but there are drawbacks to be kept in mind in the interpretation of the theoretical uncertainties obtained
using them.

First, nuclear-effect predictions based on nPDFs parametrisations significantly depend
on the factorisation scale --also referred to as Q--, $\mu_F$, at which they are evaluated. It introduces an additional 
--significant-- uncertainty which is often overlooked, whereas it is known to be already large
in the description of $pp$ collisions for which the PDFs are better known.
In \cf{fig:rpPb_scale}, we compare the $R_{\pPbm}$ obtained with 3 sets of EPS09LO with 3 choices of 
$\mu_R$, namely $(0.75,1,2) \times m_T$. As it can be seen, the effect is significant. It is essential to recall that 
no choice can be privileged. Therefore, drawing any conclusion on the strength of the nuclear
 modification of PDFs should be done with care when experimental data are only compared to a theoretical
evaluation without uncertainty on $\mu_R$.

\begin{figure}[tb!]
\subfigure[~EPS09 LO scale uncertainty]{\includegraphics[trim = 0mm 0mm 0mm 0mm, clip,height=3.5cm]{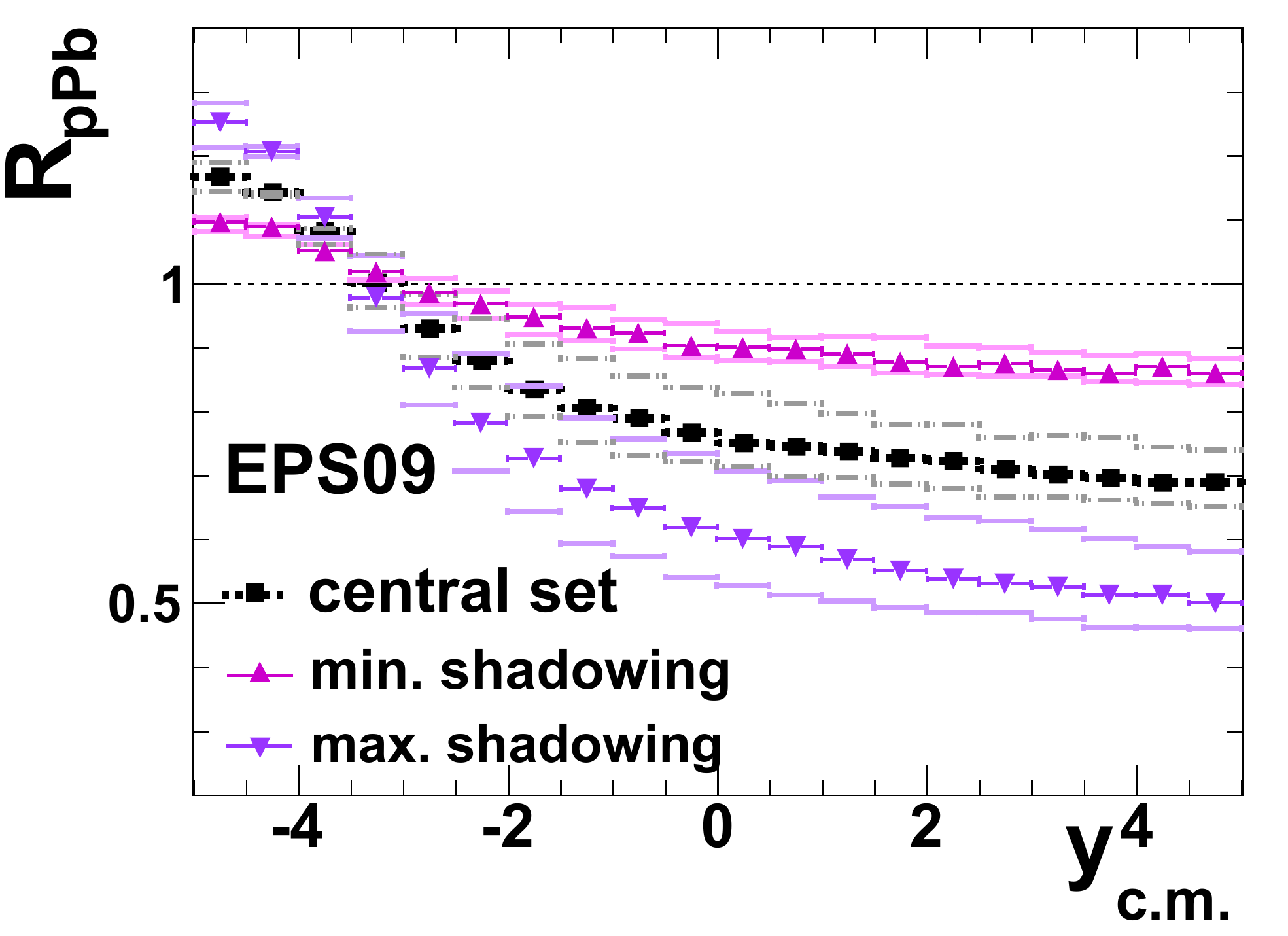}\label{fig:rpPb_scale}}\hspace*{-0.1cm}
\subfigure[~EPS09 LO vs. EPS09 NLO]{\includegraphics[trim = 30mm 0mm 0mm 0mm, clip,height=3.5cm]{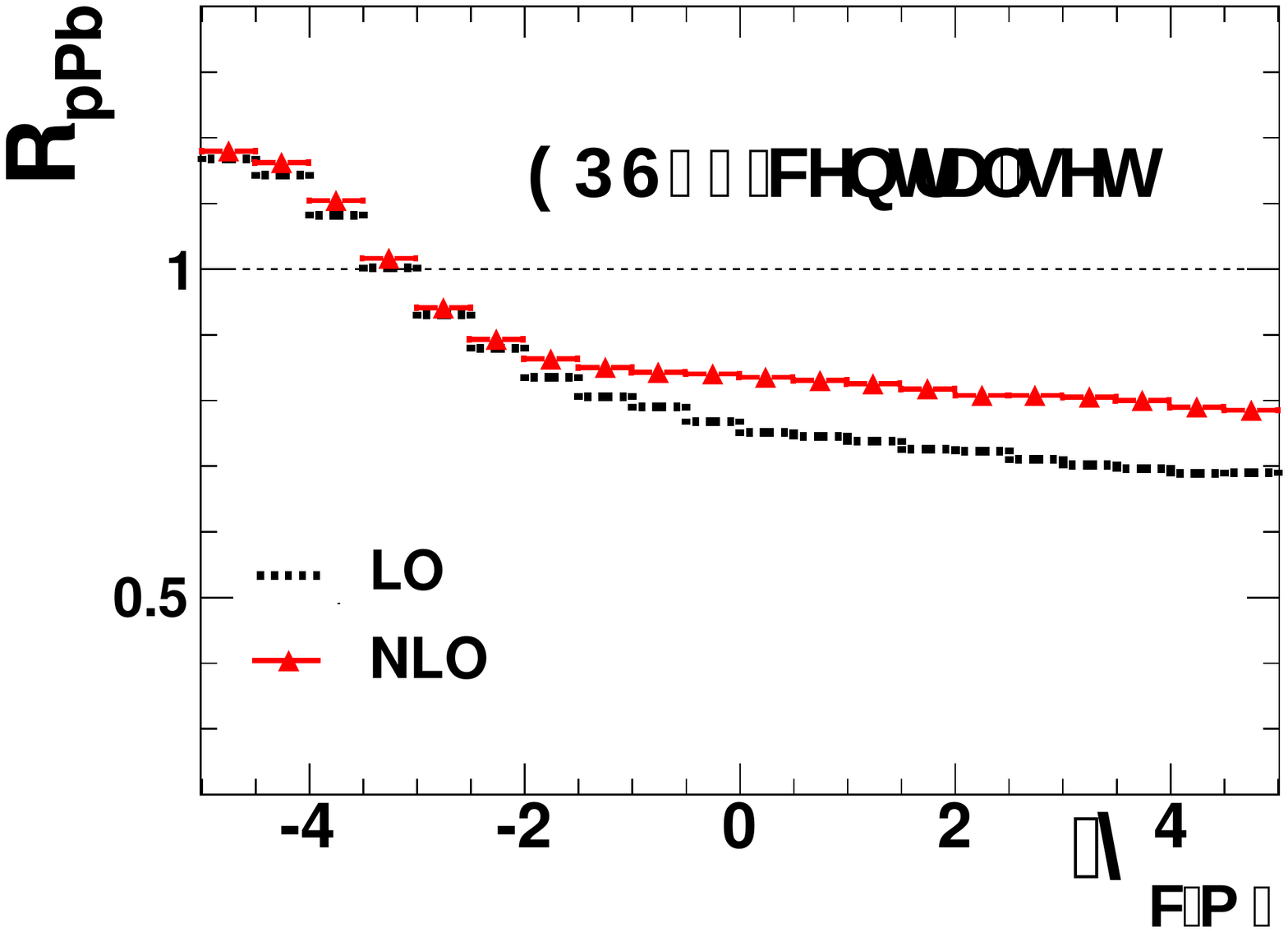}\label{fig:rpPb_LO-NLO}}
\subfigure[~EPS09 vs. nDSg]{\includegraphics[trim = 0mm 0mm 0mm 0mm, clip,height=3.5cm]{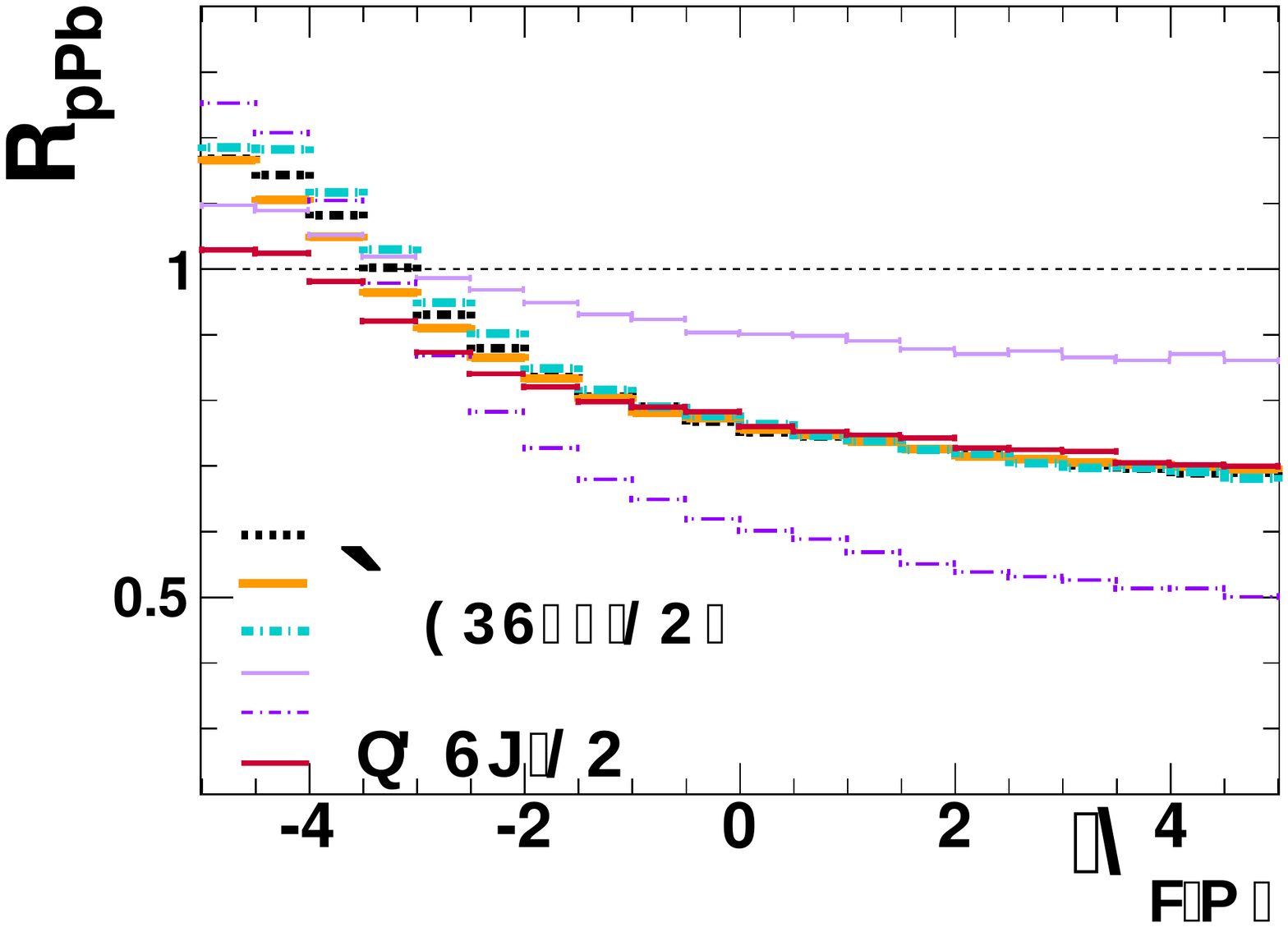}\label{fig:rpPb_nPDF}}\hspace*{-0.1cm}
\subfigure[~EPS09 LO with $\sigma_{abs}$]{\includegraphics[trim = 30mm 0mm 0mm 0mm, clip,height=3.5cm]{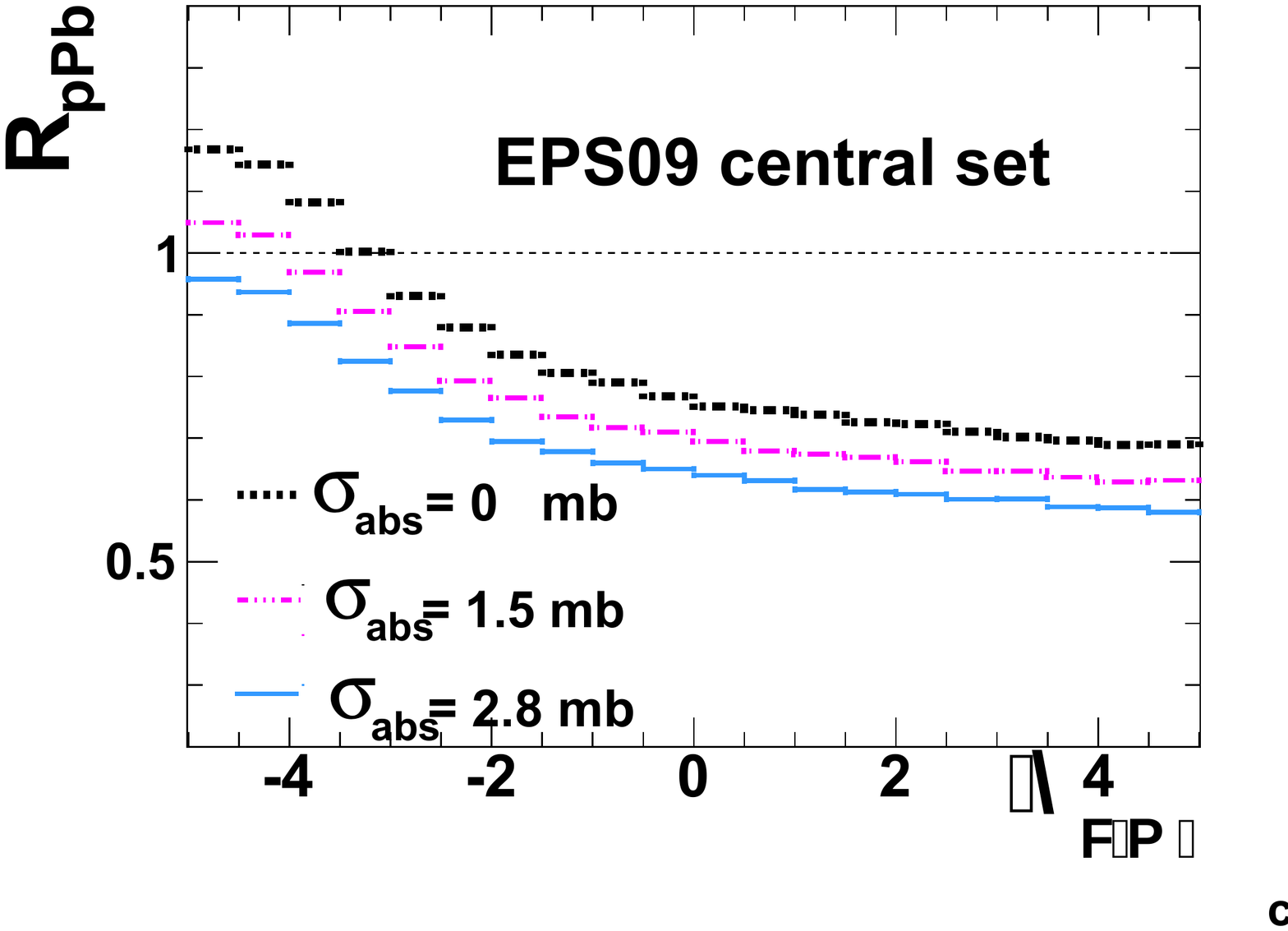}\label{fig:rpPb_abs}}
\caption{(Color online) Illustration of the uncertainties in the prediction of the  \jpsi\ nuclear modification factor in \pPb\ collisions, $R_{\pPbm}$, at $\sqrt{s_{NN}}=5\mathrm{~TeV}$ vs. $y$. (a) effect of the unknown factorisation scale taken to be $(0.75, 1, 2)\times m_T$, (b) central curves from EPS09 at LO and NLO, (c)
 extremal curves from EPS09 compared to nDSg (same color code as in Fig. 1), (d) effect of the unknown effective $c\bar c$  break up cross section for $t_f \gg R_{\rm Pb}$.}
\label{fig:rpPb_uncertainties}
\end{figure}

Second, while maybe anecdotal of EPS09, fits performed at different orders may show differences which  
may not be reflecting any specific physical phenomenon but a particular sensitivity to QCD corrections of some observables used 
in the fit at scales different than the ones used here. In our case, we are using a partonic cross section evaluated at Born (LO) 
order. The common practice is thus to employ a LO (n)PDF set. Yet, nothing forbids us to use a NLO one as a default choice. In a sense, 
any difference observed, as for instance the one between EPS09 LO and NLO on~\cf{fig:rpPb_LO-NLO}, is an indication 
of the uncertainty attributable to the neglect of unknown higher QCD corrections. 

Finally, the uncertainty spanned by a given nPDF set with error may not encompass curves which can be 
obtained by fits from different groups. This is, for instance, the case of nDSg and EPS09 LO, whose shadowing 
magnitudes are roughly the same unlike the anti-shadowing one as it can be seen on~\cf{fig:rpPb_nPDF}. Yet, 
to be rigorous, we should have derived an error band from the EPS09 eigen sets. The error band may have then
been closer to the nDSg for instance.

In addition to the uncertainties from the nPDF and their implementation, we also has to consider the effect of 
a  break-up probability on the nuclear-modification factor as in \cf{fig:rpPb_abs}.

\begin{figure}[t!]
\subfigure[~EPS09 LO scale uncertainty]{\includegraphics[trim = 0mm 0mm 1mm 0mm, clip,height=3.5cm]{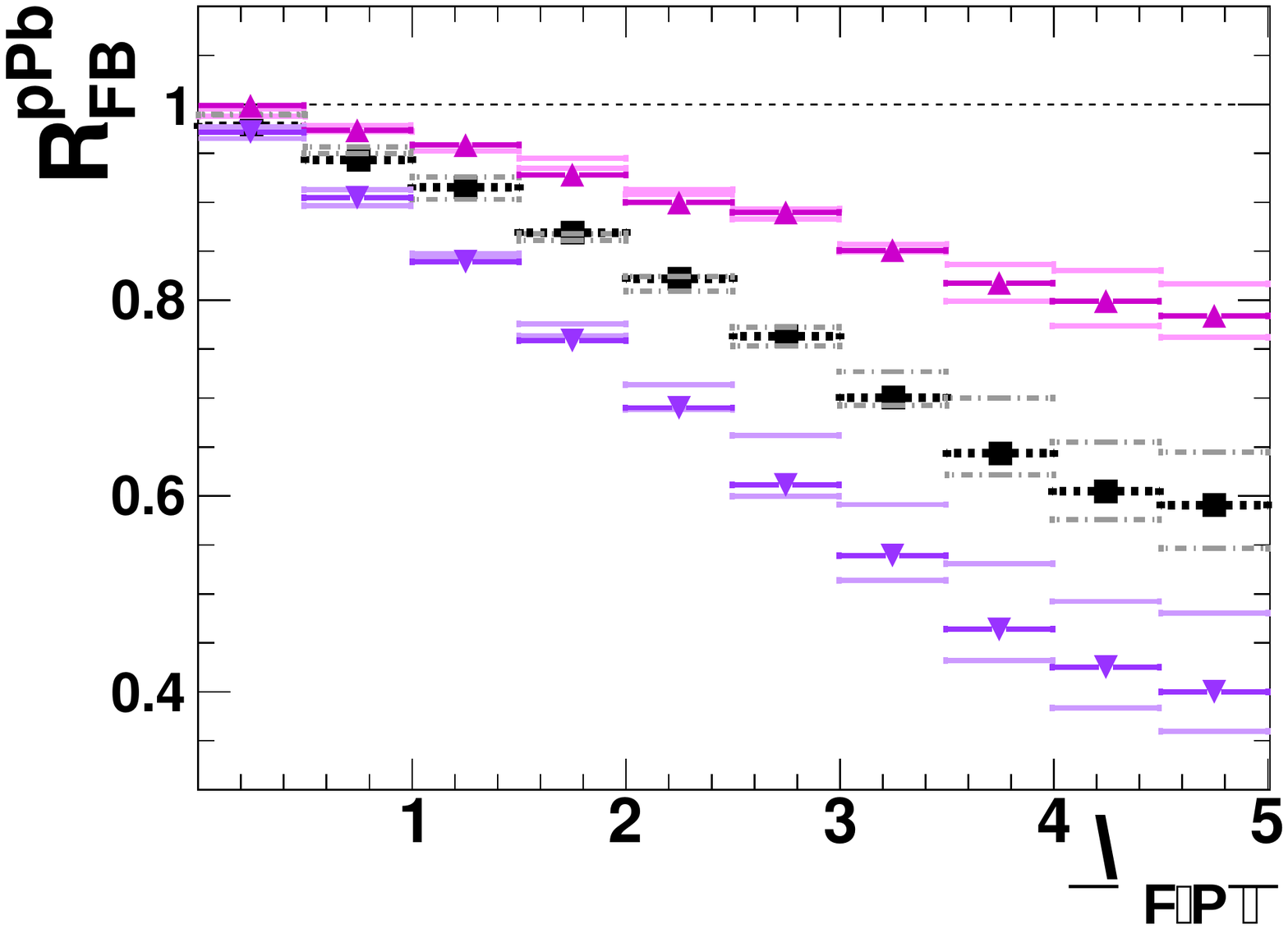}\label{fig:RFB_scale}}\hspace*{-0.1cm}
\subfigure[~EPS09 LO vs. EPS09 NLO ]{\includegraphics[trim = 29mm 0mm 0mm 0mm, clip,height=3.5cm]{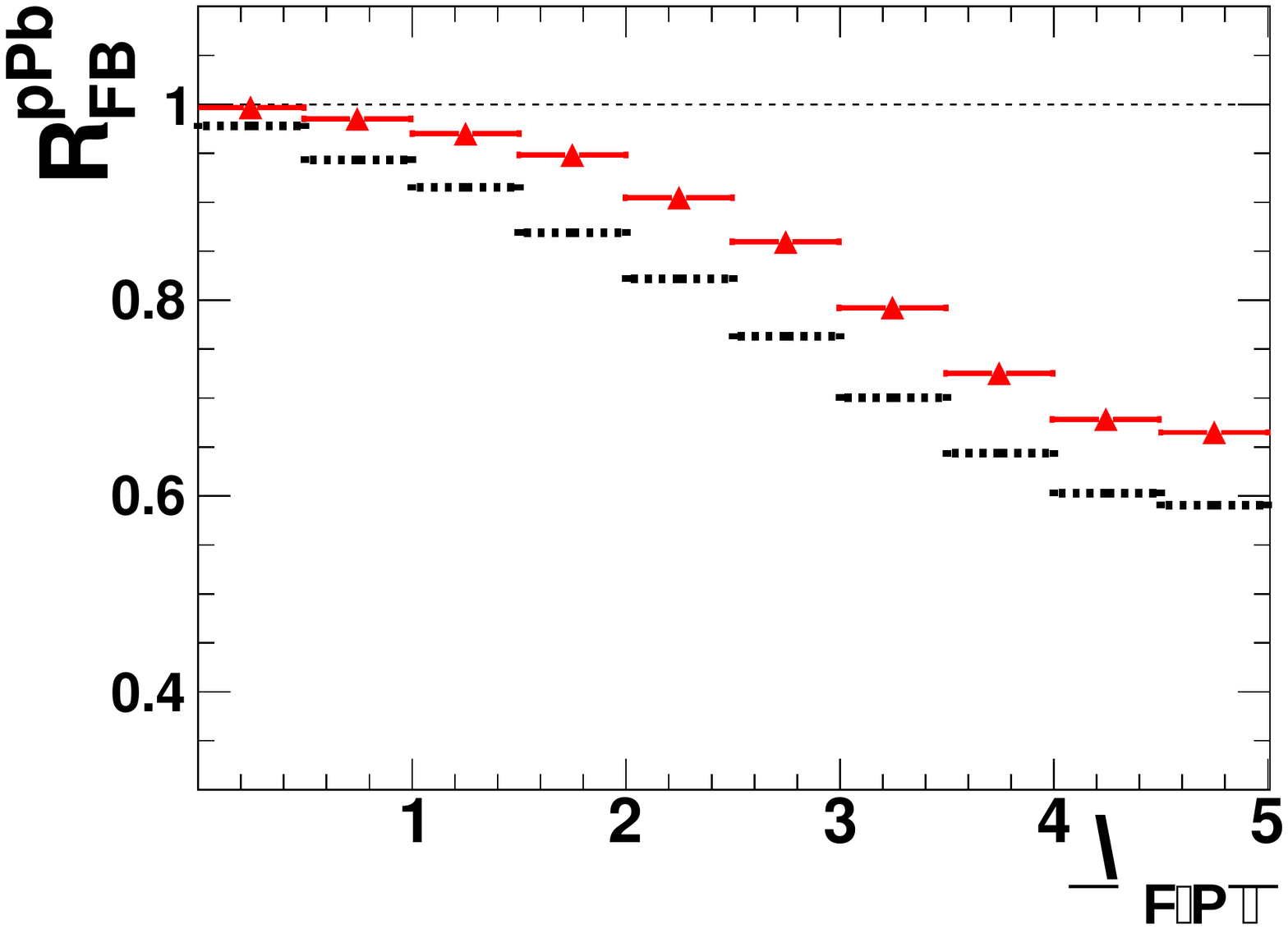}\label{fig:RFB_LO-NLO}}
\subfigure[~EPS09 vs. nDSg]{\includegraphics[trim = 0mm 0mm 1mm 0mm, clip,height=3.5cm]{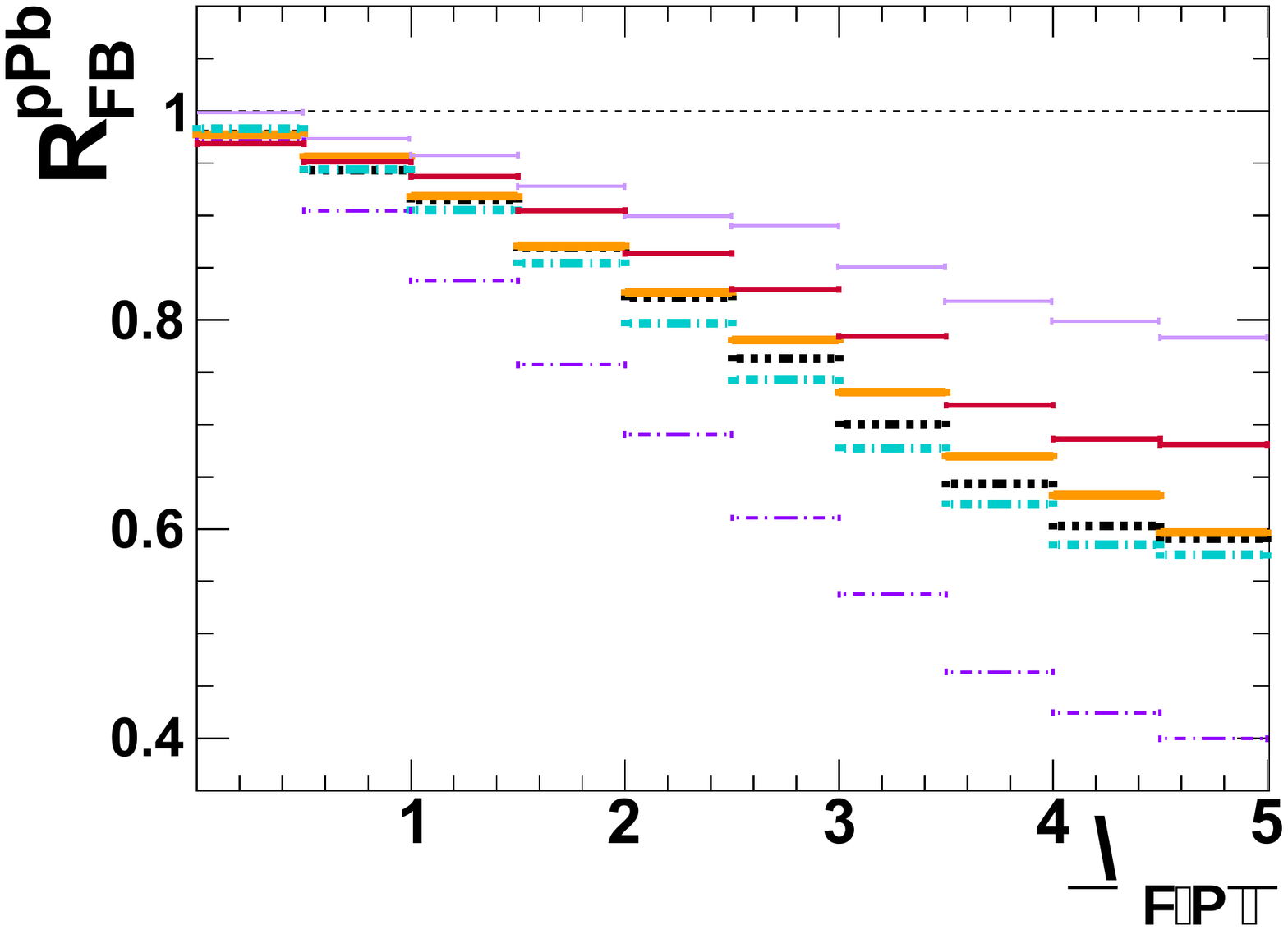}\label{fig:RFB_nPDF}}\hspace*{-0.1cm}
\subfigure[~EPS09 LO with $\sigma_{abs}$]{\includegraphics[trim = 29mm 0mm 0mm 0mm, clip,height=3.5cm]{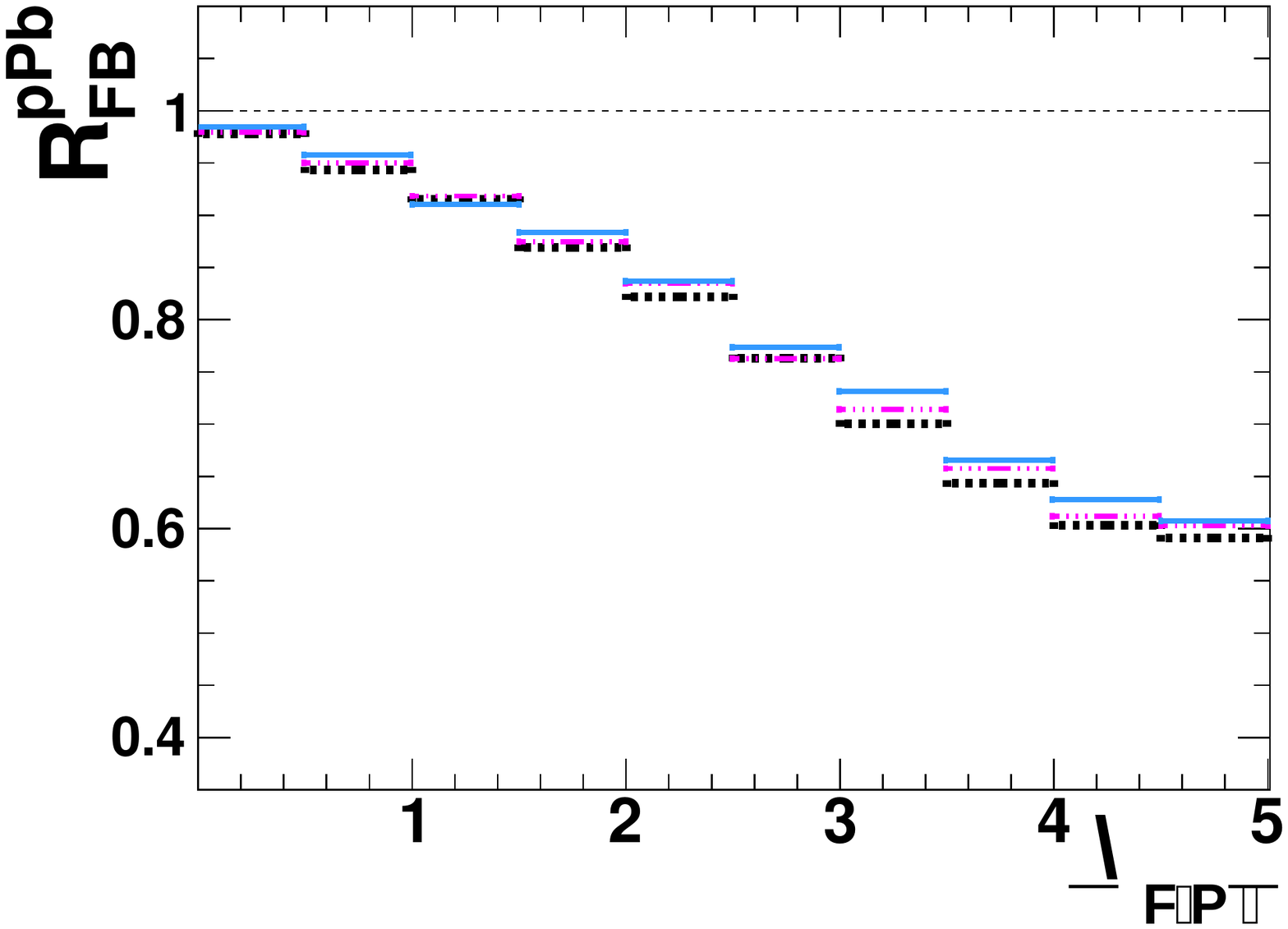}\label{fig:RFB_abs}}
\caption{(Color online) Same as \cf{fig:rpPb_uncertainties} for $R_{\rm FB}$}\vspace*{-.5cm}
\label{fig:RFB_uncertainties}
\end{figure}

All these effects also impact the forward-to-backward ratio $R_{\rm FB}$ as can be seen on \cf{fig:RFB_uncertainties}. 
Their impact remains significant, except for the break-up probability. Its 
effect on $R_{\rm FB}$ (see \cf{fig:RFB_abs}) is much smaller than on \RpPb\ and can in practice be disregarded.

These values can be compared to the preliminary measurements by the LHCb and ALICE collaborations.
LHCb has reported~\cite{LHCb-CONF-2013-008} for the prompt $J/\psi$ yield a preliminary value of
$R_{\rm FB} = 0.66 \pm 0.08$ for $2.5 < |y_{\rm CM}| < 4$, whereas ALICE has  reported~\cite{ALICE} 
 a preliminary value of $R_{\rm FB} = 0.60 \pm 0.07$ for $3 < |y_{\rm CM}| < 3.5$ for the inclusive one.
These values are compatible with a strong shadowing.

We would like to recall that in order to take into account the transverse-momentum dependence of the shadowing 
effects one needs to resort to a model which contains an explicit dependence on $P_T$ and $y$. 
Thanks to the versatility of our Glauber code, such computation can also be done including
the impact-parameter dependence along with involved production mechanisms containing 
a non-trivial $P_T$-dependence.   

\begin{figure}[H]
\begin{center}
\subfigure[~$R_{\pPbm}$ for $P_T~>~6$ GeV: EPS09LO \newline vs. nDSg]
{\includegraphics[trim = 0mm 0mm 3mm 0mm, clip,height=3.4cm]{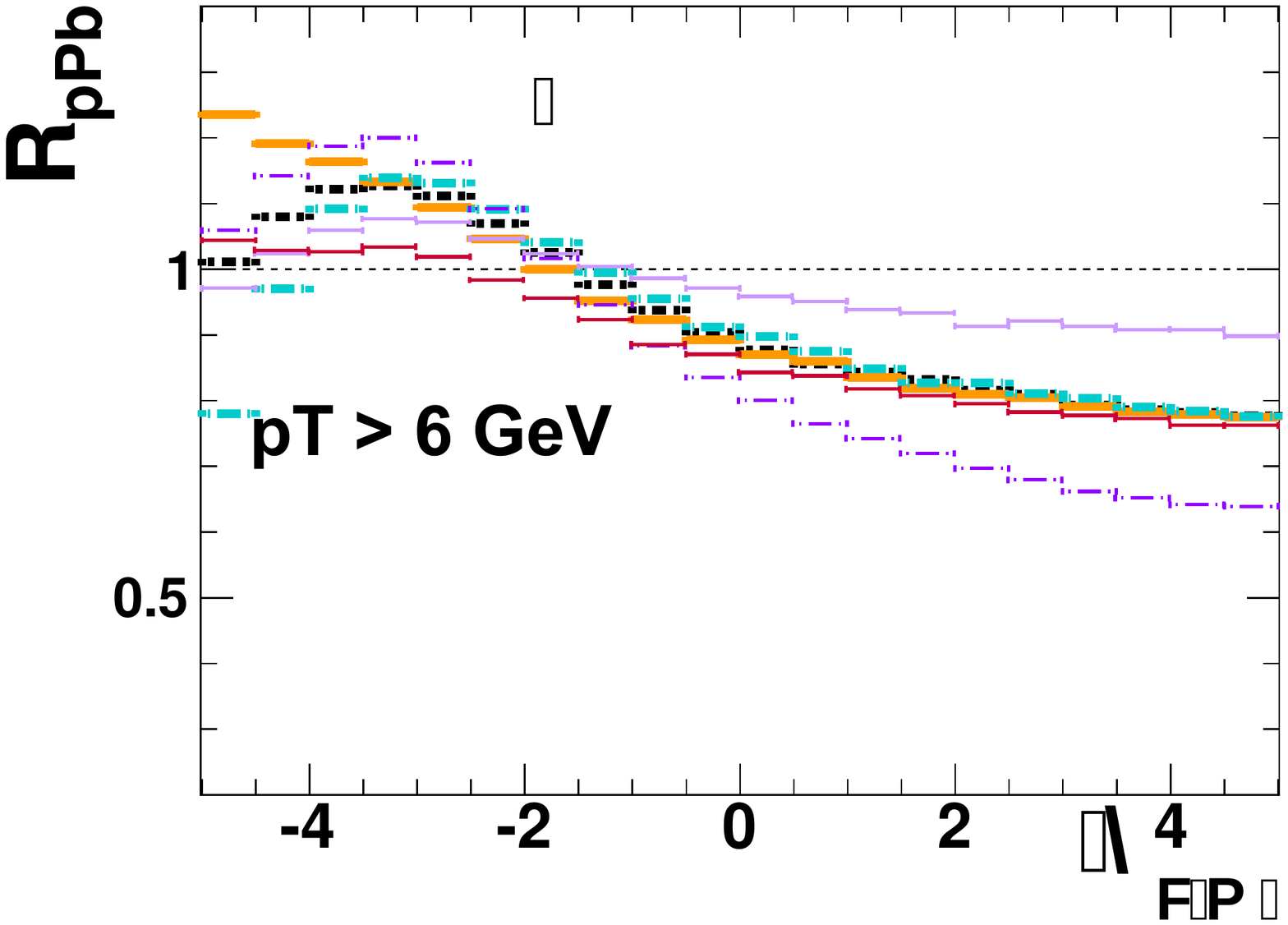}\label{fig:rpPb-ptcut}}
\hspace*{-0.2cm}
\subfigure[~$R_{\pPbm}$ scale uncertainty  \newline for $P_T~>~6$ GeV]{\includegraphics[trim = 29mm 0mm 0mm 0mm, clip,height=3.4cm]{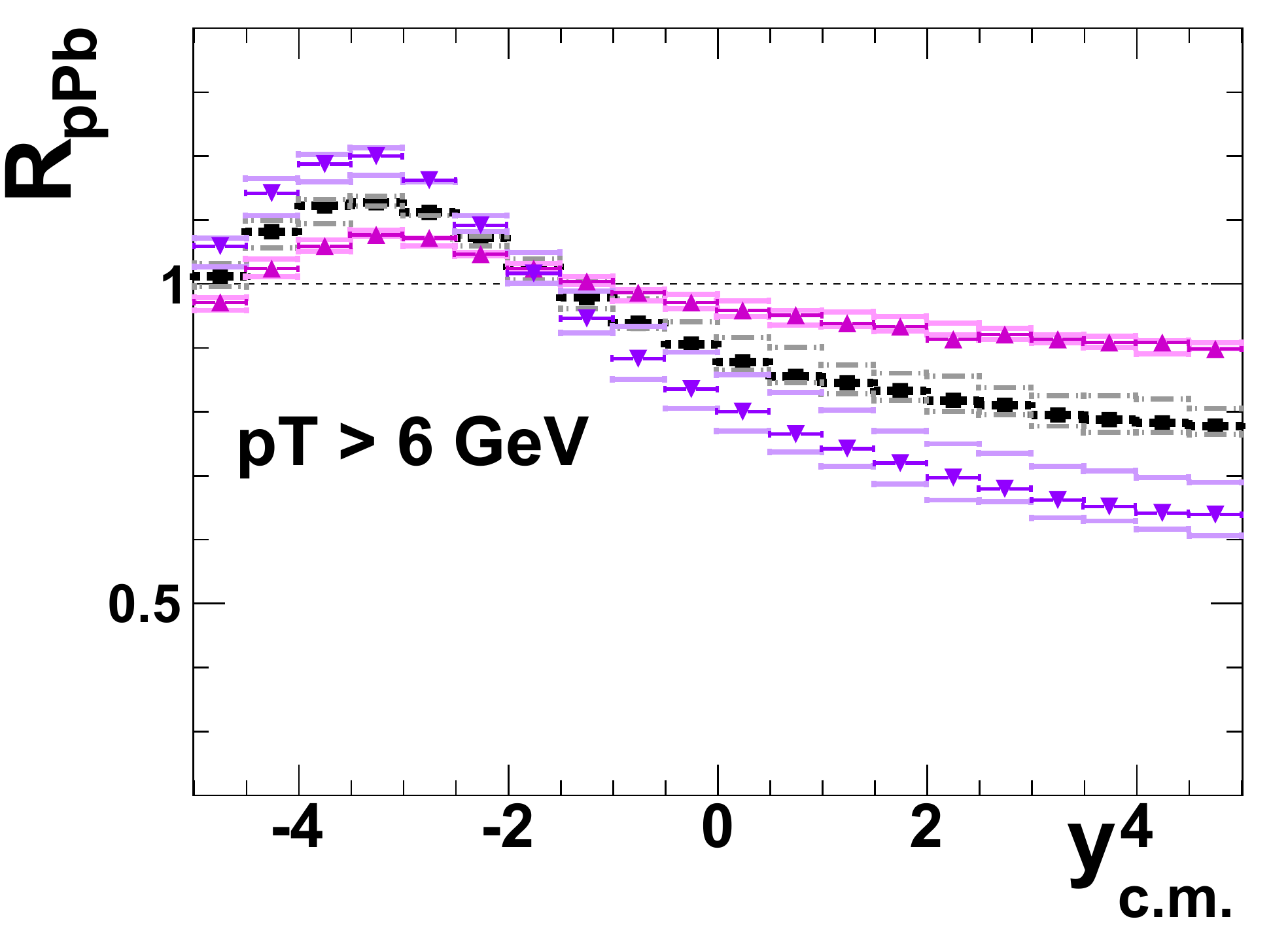}\label{fig:rpPb_scale-ptcut}}\vspace*{-0.3cm}\\
\subfigure[~$R_{\rm FB}$ for $P_T~>~6$ GeV: EPS09LO \newline  vs. nDSg ]
{\includegraphics[trim = 0mm 0mm 3mm 0mm, clip,height=3.4cm]{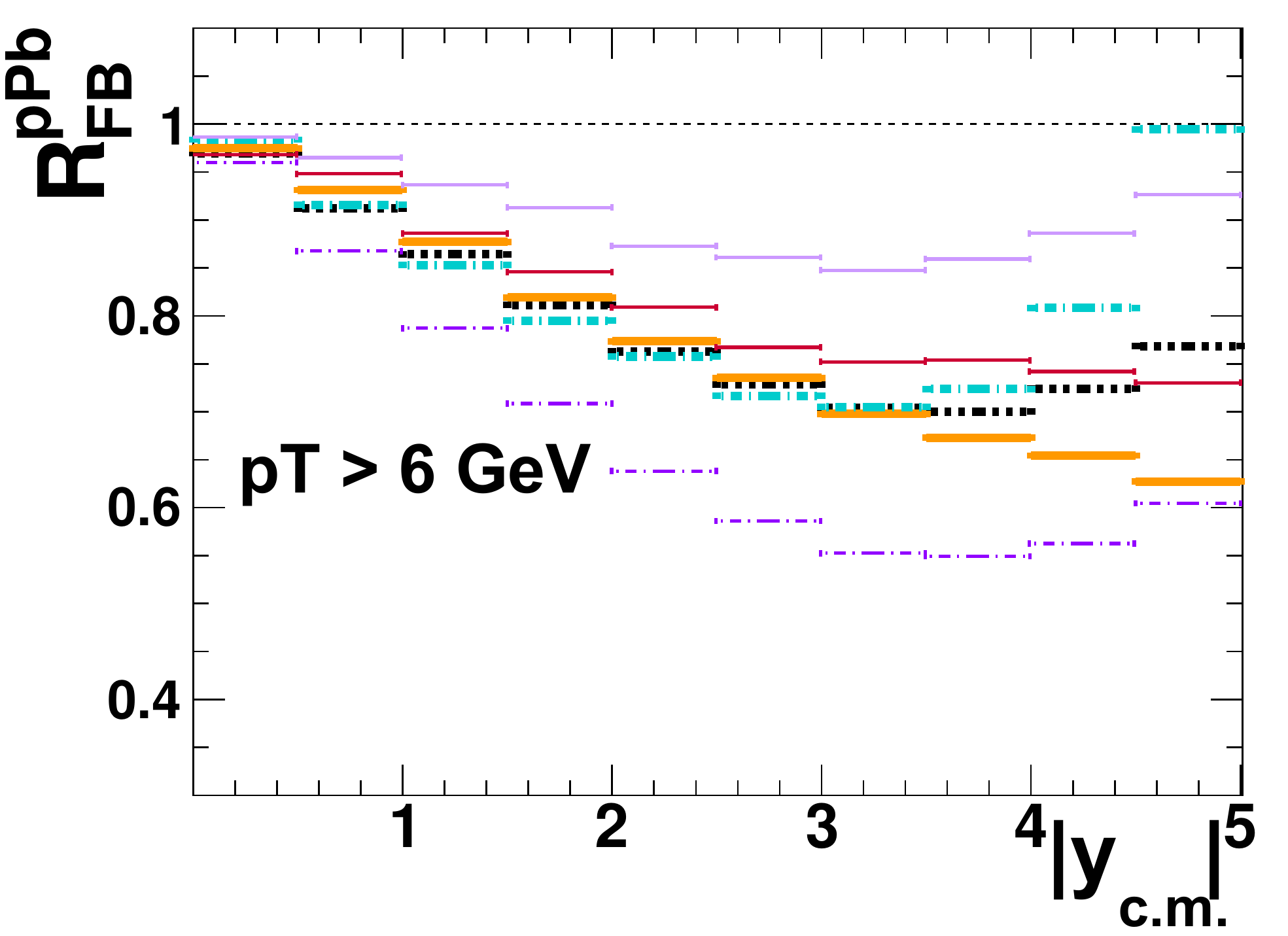}\label{fig:RFB-ptcut}}
\hspace*{-0.2cm}
\subfigure[~$R_{\rm FB}$ scale uncertainty \newline for $P_T~>~6$ GeV ]{\includegraphics[trim = 29mm 0mm 0mm 0mm, clip,height=3.4cm]{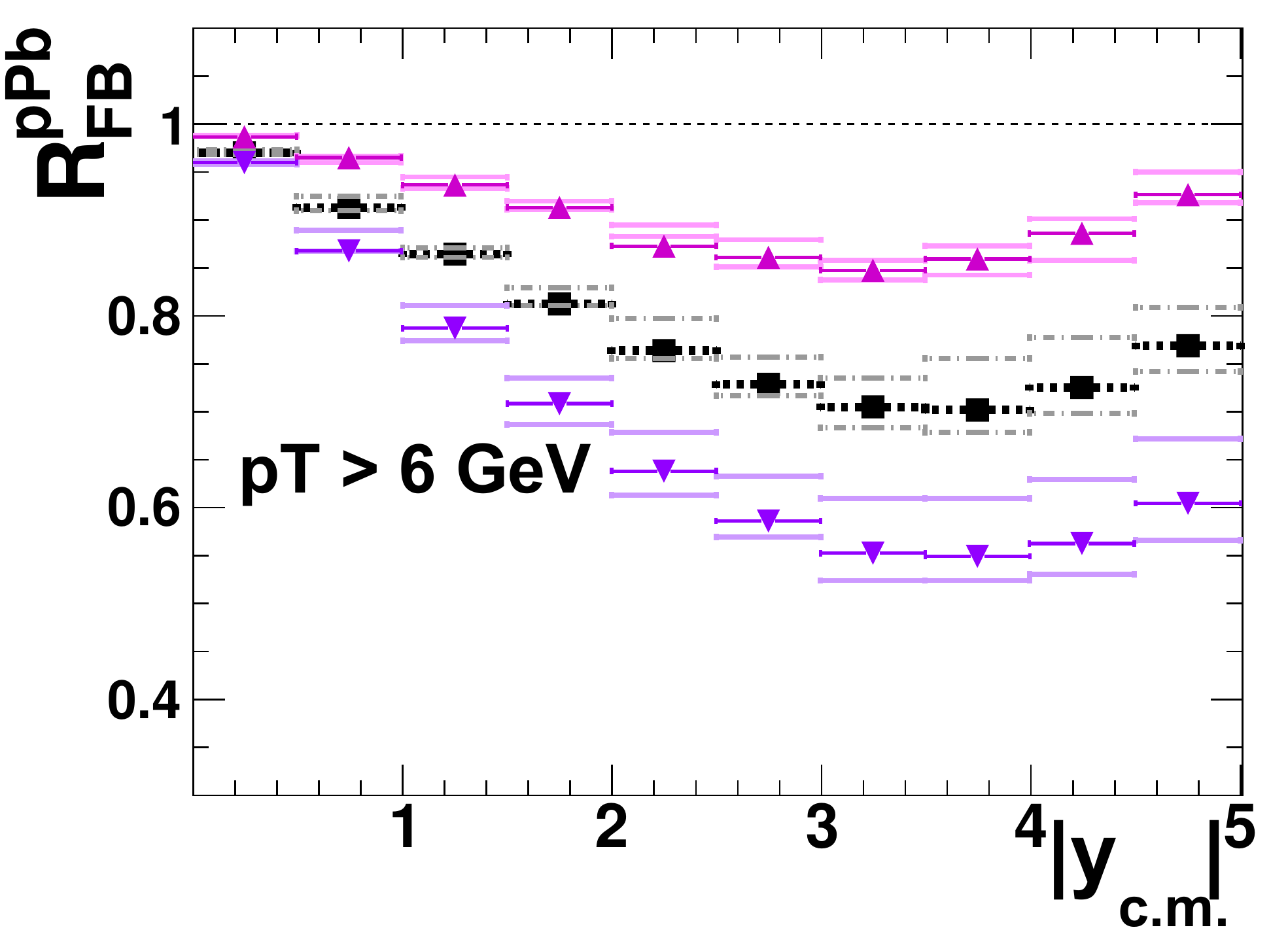}\label{fig:RFB_scale-ptcut}}\vspace*{-0.3cm}
\end{center}
\caption{(Color online) Effect of the cut $P_T > 6$ GeV on the \jpsi\ nuclear modification factor (a \& b) 
and its forward-to-backward ratio (c \& d) in \pPb\ collisions 
 at $\sqrt{s_{NN}}=5\mathrm{~TeV}$ versus $y$.
}\label{fig:ptcut}\vspace*{-0.4cm}
\end{figure}

The effect of imposing a $P_T$ cut is shown  on \cf{fig:ptcut}. We emphasise that it is not related to any Cronin effect 
which is not taken into account here.
It simply comes from the increase of $x_2$  and $m_T(P_T)$ for increasing $P_T$. In particular, the anti-shadowing peak is shifted to less
negative rapidities which modifies $R_{\rm FB}$ at large $|y_{CM}|$ (\cf{fig:RFB-ptcut}). The uncertainty coming from the choice of the factorisation 
scale is not much reduced (\cf{fig:rpPb_scale-ptcut} \& \ref{fig:RFB_scale-ptcut})  despite the larger value of $m_T$.

\begin{figure}[htb!]
\includegraphics[trim = 0mm 0mm 0mm 0mm, clip,width=\columnwidth]{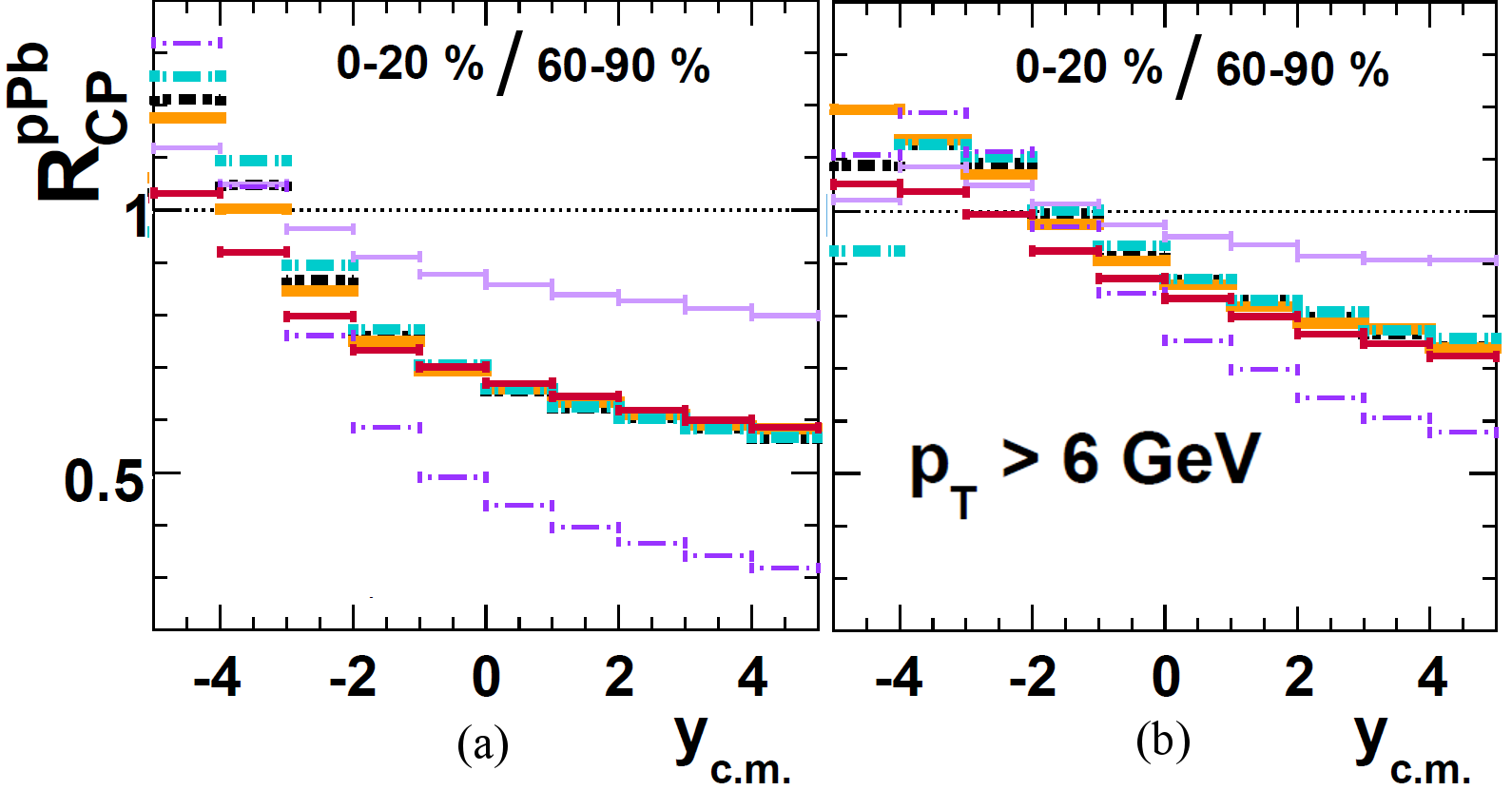}
\caption{(Color online) Central-over-peripheral ratio for 5 sets of EPS09 and nDSg as function of $y$ without (a) and with (b) a $P_T$ cut.}
\label{fig:R_CP}\vspace*{-.25cm}
\end{figure}

Finally, we show on \cf{fig:R_CP} our predictions for $R_{\rm CP}$ for the 5 sets of EPS09 which we used and nDSg 
with and without a $P_T$ cut. We see that the trend of $R_{\rm CP}$ is similar to that of $R_{\pPbm}$, 
with a larger magnitude. $R_{\pPbm}$ in four rapidity classes are provided as supplemental material in appendix~\ref{SM}.
Since part of the experimental uncertainties and that related to the unknown $pp$ reference cancel in $R_{\rm CP}$, we are hopeful that
forthcoming data will tell us more about the magnitude of the gluon shadowing despite the theoretical complications which we enumerated before.

{\it Conclusions.---}
We have provided predictions for the \jpsi\ nuclear  modification factors in \pPb\ collisions, its  forward-to-backward ratio, $R_{\rm FB}$
and its central-over-peripheral ratio, $R_{\rm CP}$,   at $\sqrt{s_{NN}}=5\mathrm{~TeV}$
as functions of $y$ for low and mid $P_T$, which can be compared to the ALICE and LHCb preliminary data
and the forthcoming ATLAS and CMS one taken during the 2013 LHC  \pPb\ run.

Unless there is an unexpectedly large anti-shadowing, the measured values of $R_{\rm FB}$ support the presence of
a significant shadowing --of a magnitude stronger than the  central value of EPS09LO. 
This should be taken into account in the interpretation the $J/\psi$ PbPb data.


{\small We would like to thank R. Arnaldi, Z. Conesa del Valle, K. Eskola, C.~Hadjidakis, J. He, F. Jing, N.~Matagne, 
I. Schienbein, R.~Vogt and Z. Yang   for stimulating and useful discussions.
}



\newpage 
\appendix 

\begin{widetext}
\section{Supplemental material}\label{SM}

\begin{center}
4  plots of $R_{\pPbm}$ in four rapidity classes without $P_T$ cut\\
\includegraphics[width=11cm]{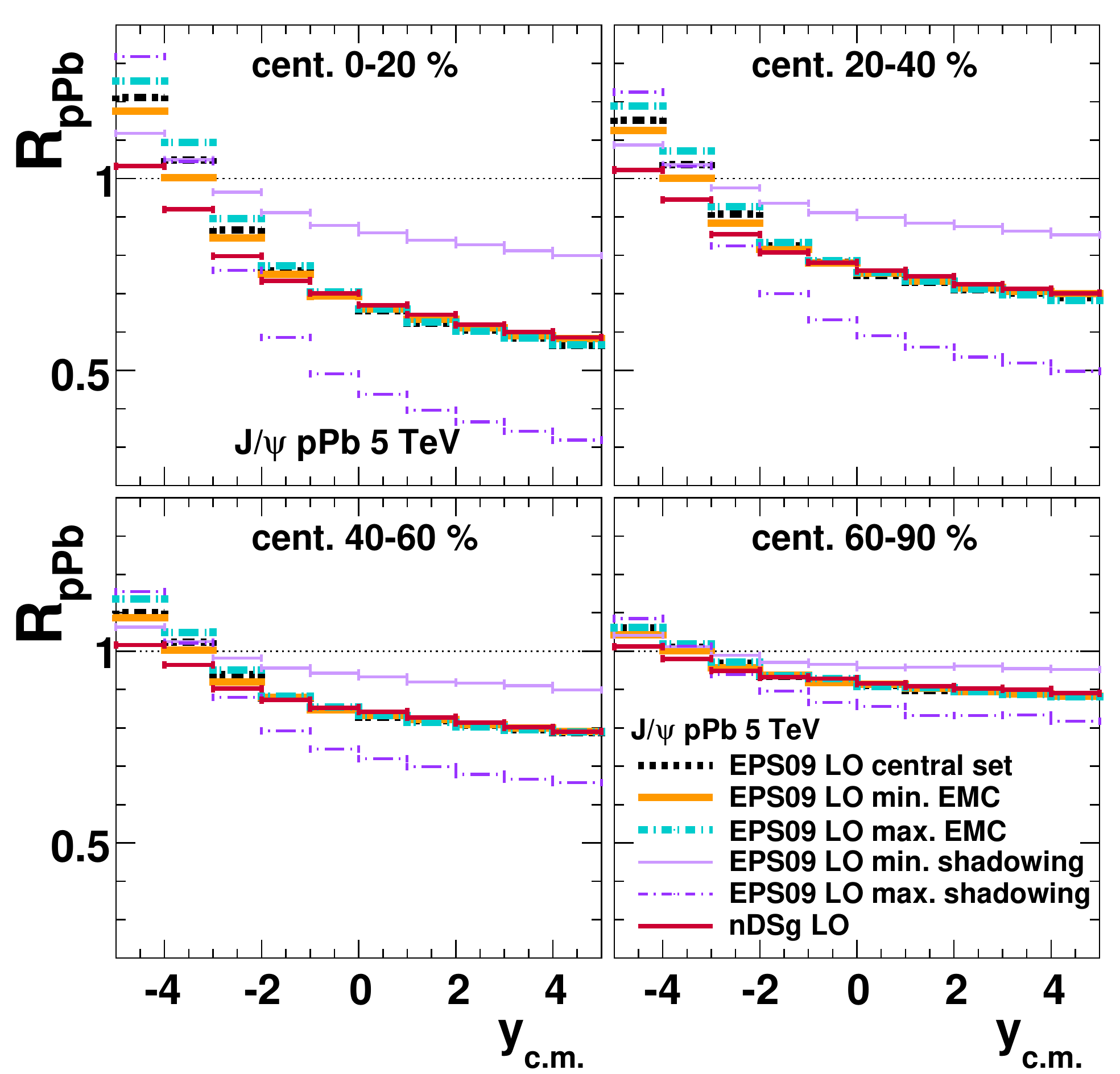}\\
4  plots of $R_{\pPbm}$ in four rapidity classes with $P_T$ cut\\
\includegraphics[width=11cm]{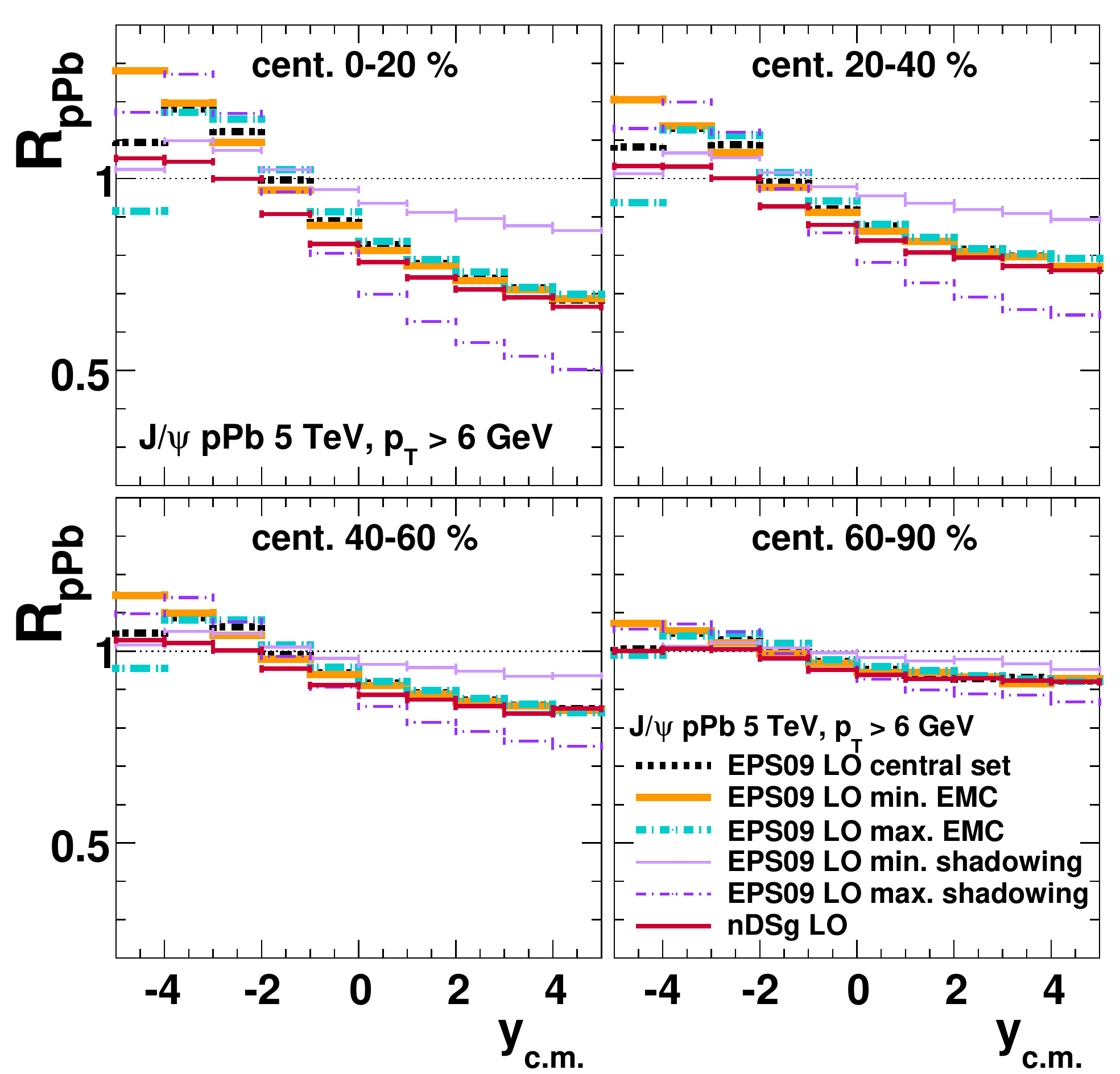}
\end{center}

\end{widetext}
\end{document}